%% file: main.tex
\documentclass[runningheads]{llncs}
\usepackage[T1]{fontenc}
\usepackage{graphicx}
\usepackage{cite}
\usepackage{amsmath,amssymb,amsfonts}
\usepackage{algorithmic}
\usepackage{graphicx}
\usepackage{textcomp}
\usepackage{xcolor}
\usepackage{booktabs}
\usepackage{listings}

\usepackage{tikz}
\usepackage{amsmath}
\usepackage{textcomp}
\usepackage{bmpsize}
\usepackage{xcolor}
\usepackage{lipsum}
\usepackage[ruled,linesnumbered]{algorithm2e}
\usepackage{dblfloatfix}
\usepackage[english]{babel}
\usepackage{multirow}
\usepackage{amsmath,array,graphicx}
\usepackage{todonotes}
\usepackage{float}
\usepackage{hyperref}

\newcommand{\cmmnt}[1]{}
\usepackage{amsmath,amsfonts}
\usepackage{booktabs}
\newcommand{\set}[1]{\ensuremath{\mathcal{#1}}}
\usepackage{makecell} 

\newcommand\Tstrut{\rule{0pt}{5ex}} 

\usepackage{array} 

\usepackage{orcidlink}
\usepackage{amsfonts}

\usepackage{eso-pic}
\usepackage{lipsum} 

\AddToShipoutPictureBG*{%
  \AtPageUpperLeft{%
    \setlength\unitlength{1in}%
    \raisebox{-2cm}[0pt][0pt]{
      \makebox[0pt][l]{\hspace*{4.7cm}\parbox{\textwidth}{ \fontsize{8}{11}\selectfont Accepted to the ESORICS 2024 Workshop on Security and Artificial Intelligence (SECAI 2024)\vspace{-0.3cm}\\\rule{\textwidth}{0.4pt}}}%
    }%
  }%
}

\begin{document}

\title{Improving Adversarial Robustness in Android Malware Detection by Reducing the Impact of Spurious Correlations}
\titlerunning{Improving Adversarial Robustness in Android Malware Detection}


%
%
\author{Hamid Bostani\inst{1}\orcidlink{0000-0002-2097-5521} \and
Zhengyu Zhao\inst{2}\orcidlink{0000-0003-0745-4294} \and
Veelasha Moonsamy\inst{3}\orcidlink{0000-0001-6296-2182}}
\authorrunning{H. Bostani et al.}
%
\institute{Digital Security Group, Institute for Computing and Information Sciences, Radboud University, Nijmegen, The Netherlands
\email{hamid.bostani@ru.nl}\\
\and
Faculty of Electronic and Information Engineering, Xi'an Jiaotong University, Xi'an, China
\email{zhengyu.zhao@xjtu.edu.cn}
\and
Horst Görtz Institute for IT Security, Ruhr University Bochum, Bochum, Germany\\
\email{email@veelasha.org}}

\maketitle              
\begin{abstract}

Machine learning (ML) has demonstrated significant advancements in Android malware detection (AMD); however, the resilience of ML against realistic evasion attacks remains a major obstacle for AMD. One of the primary factors contributing to this challenge is the scarcity of reliable generalizations. Malware classifiers with limited generalizability tend to overfit spurious correlations derived from biased features. Consequently, adversarial examples (AEs), generated by evasion attacks, can modify these features to evade detection. In this study, we propose a domain adaptation technique to improve the generalizability of AMD by aligning the distribution of malware samples and AEs. Specifically, we utilize meaningful feature dependencies, reflecting domain constraints in the feature space, to establish a robust feature space. Training on the proposed robust feature space enables malware classifiers to learn from predefined patterns associated with app functionality rather than from individual features. This approach helps mitigate spurious correlations inherent in the initial feature space. Our experiments conducted on DREBIN, a renowned Android malware detector, demonstrate that our approach surpasses the state-of-the-art defense, Sec-SVM, when facing realistic evasion attacks. In particular, our defense can improve adversarial robustness by up to 55\% against realistic evasion attacks compared to Sec-SVM.

\end{abstract}
\input{01-introduction}
\input{02-background}
\input{03-our_proposed_defense}
\input{04-experiments}
\input{05-related_work}
\input{06-limitations_and_future_work}

\input{07-conclusion}
\subsubsection*{Acknowledgements.} 
Veelasha Moonsamy was supported by the Deutsche Forschungsgemeinschaft (DFG, German Research Foundation) under Germany’s Excellence Strategy - EXC 2092 CASA - 390781972.
\bibliographystyle{unsrt}
\bibliography{bibliography}
\end{document}

%% file: 01-introduction.tex
\section{Introduction}
\label{section:introduction}
Despite the substantial progress in utilizing machine learning (ML) for Android malware detection (AMD), the field still suffers from security concerns surrounding ML models, especially their vulnerabilities to realistic evasion attacks. These attacks change malware apps into adversarial examples (AEs), tricking AMD while preserving the malware's properties, e.g., their executability and malicious functionalities. Evasion attacks can circumvent ML models, exploiting their susceptibility to learning vulnerabilities~\cite{maiorca2020adversarial}, which often arise from the limited generalizability inherent in ML models. In fact, adversaries deceive ML models by generating AEs that sufficiently deviate from the distribution of the training samples~\cite{zhang2019limitations}. 

One of the major factors contributing to the generalizability challenges of ML is \textit{biased features}\cite{he-etal-2019-unlearn}. These features introduce biases into the model, potentially resulting in poor performance on unseen samples. ML models tend to learn these simple cues effectively on most training samples, which in turn causes them to perform well on those samples; however, they encounter difficulty with more complex unseen samples (e.g., AEs) due to the distribution shift~\cite{he-etal-2019-unlearn}. 
Specifically, the presence of biased features causes classifiers to learn \textit{spurious correlations}~\cite{ye2024spurious}, resulting in misleading associations between biased features and the target variable, known as the \textit{label} in supervised learning. In other words, ML models might learn misleading patterns--the irrelevant associations between features and the label--that may not generalize well to unseen samples with distributions different from those of the training samples. These misleading correlations often occur because the training set fails to accurately represent the true data distribution, typically due to sampling bias~\cite{cortes2008sample}. Spurious correlations, substantially diminish the generalizability of ML models, especially in the context of cybersecurity. These meaningless correlations represent patterns within the data unrelated to the security problem but serve as shortcuts for distinguishing classes~\cite{quiring2022and}. For instance, the inclusion of particular market information, such as Chinese markets, in numerous malware samples might cause the ML model to mistakenly associate this feature with maliciousness~\cite{quiring2022and}, instead of prioritizing the identification of authentic patterns linked to malicious behavior.

\begin{figure}[!t]
    \centering
    \includegraphics[width=0.9\columnwidth]{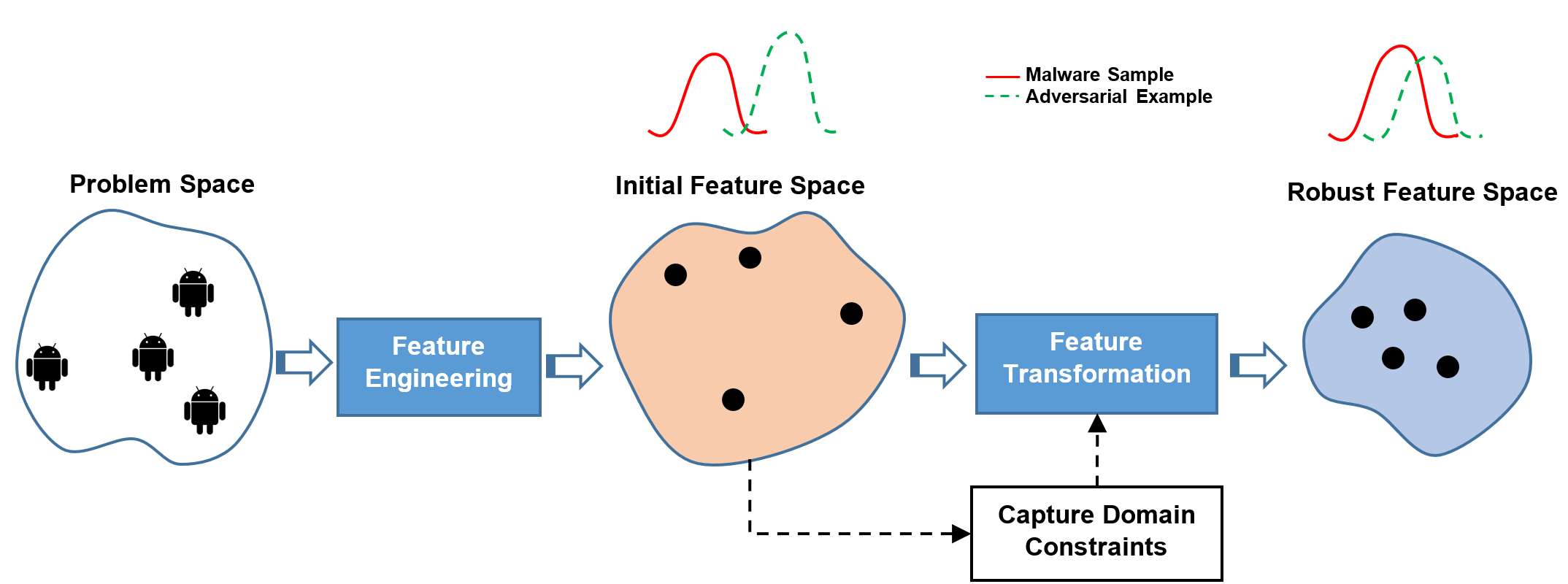}
    \caption{An illustration of our proposed domain-adaptation technique. In the initial feature space, the distributions of malware samples and adversarial examples differ significantly. However, in our proposed robust feature space, their distributions are more aligned.}
    \label{fig:overview}
\end{figure}

Spurious correlations present intriguing implications for realistic evasion attacks. Adversaries aim to append adversarial payloads that significantly influence features crucial for classification while ensuring that the added contents are unrelated to the app's functionality.
Therefore, realistic evasion attacks could leverage the features associated with spurious correlations, as they influence the classification outcome while ensuring the malicious patterns remain intact. Since learning spurious correlations poses challenges for malware classifiers when the distribution of AEs significantly differs from that of malware samples in the training set, this issue can be reduced if both malware samples and AEs follow a similar distribution. This study introduces a novel domain adaptation\footnote {In domain adaptation~\cite{ben2006analysis}, \textit{domain} refers to a certain distribution over a sample set (e.g., the training set).} technique designed to reduce the impact of spurious correlations by aligning the distributions of the source domain (including malware samples) and the target domain (including AEs). As illustrated in Figure~\ref{fig:overview}, to reduce the adverse impact of spurious correlations on the adversarial robustness of AMD, our proposed approach utilizes domain constraints that characterize app properties to create a robust feature space. To this end, we first model domain constraints based on the relationships between features derived from the feature representations of the training apps. Within the feature space, domain constraints denote complex relationships among features that an adversary must fulfill for an attack to be realistic~\cite{b51}. Then, we propose a transformation function using these identified patterns to transform samples from the initial feature space to a robust feature space.
The distribution of malware apps is expected to align more closely with adversarial ones when represented in the robust feature space rather than in the initial feature space. This is because, unlike the features in the initial feature space, each feature in the robust space reflects a functional aspect (e.g., sending an SMS) commonly shared among feasible apps, including both malware and adversarial apps. Our contributions\footnote{We make our code publicly available at {\url{https://github.com/HamidBostani2021/robust-feature-space}} to allow reproducibility.} can be summarized as follows:
\begin{itemize}
\item We propose a robust feature space based on a novel domain adaptation approach to reduce spurious correlations, thereby enhancing the adversarial robustness of AMD against realistic evasion attacks. 
\item We empirically demonstrate that our proposed defense surpasses the state-of-the-art defense, Sec-SVM~\cite{b19}, in hardening AMD against gradient-based and query-based realistic attacks across various threat models.
\item Our empirical findings illustrate that the distribution of malware apps and AEs is more aligned in the proposed robust feature space than in the initial feature space. 
\end{itemize}

The rest of the paper is organized as follows: Section~\ref{section:background} provides an elaboration on the fundamental concepts crucial to the paper. Our novel approach for constructing a robust feature space is detailed in Section~\ref{section:proposed_defense}. Section~\ref{section:experiments} assesses the effectiveness of our proposed defense technique in hardening the robustness of ML-based AMD against realistic evasion attacks. Section~\ref{section:related-work} examines relevant studies that have explored feature representations to enhance the adversarial robustness of malware detection. The limitations and potential directions for future research are discussed in Section~\ref{sec:limitations_and_future_work} and we conclude in Section~\ref{section:conclusion}.

%% file: 02-background.tex
\section{Background}
\label{section:background}
In this section, we briefly review the fundamentals of evasion attacks, spurious correlations, and domain constraints.
\subsection{Evasion Attacks}
\label{subsec:evasion_attacks}
Consider $\phi: \mathcal{Z} \rightarrow \mathcal{X}$ as a mapping function representing Android apps of the problem space $\mathcal{Z}$ with $d$-dimensional feature vectors in the feature space $\mathcal{X}$. ML-based AMD is a binary classifier $f: \mathcal{X} \rightarrow \mathcal{Y}$ equipped with a discriminant function $g: \mathcal{X} \times \mathcal{Y} \rightarrow \mathbb{R}$. Here, $f(x) = \arg\max_{i \in \mathcal{Y}} g_i(x)$ assigns labels to $x \in \mathcal{X}$, where $\mathcal{Y} = \{0,1\}$ denotes the label space with $y = 0$ representing benign labels and $y = 1$ representing malicious labels. In the binary feature space~\cite{b14, b15, b35}, each element of the feature vector $x \in \mathcal{X}$ is binary, where $0$ signifies absence and $1$ signifies the presence of specific features. It is noteworthy that $\set F = \{f_1, f_2, ..., f_d\}$ represents the feature set defining the dimensions of $\mathcal{X}$, where $d$ denotes the number of dimensions.

Generally, evasion attacks generate AEs by altering $x \in \mathcal{X}$ through the discovery of optimal perturbations $\delta$ applied to it. Particularly, malicious actors endeavor to solve the following optimization problem~\cite{b17, b19}:

\begin{equation} \label{eq_optimization_feature_space}
\begin{aligned}
\arg\max_{\delta} \quad & g_{0}(x' = x + \delta) \quad \text{subject to} \quad \delta \models \Omega,
\end{aligned}
\end{equation}

where the perturbation vector $\delta$ must satisfy constraints defined in the feature space denoted by $\Omega$, such as a naive norm bound~\cite{b17}.

\subsection{Spurious Correlations}
In statistical analysis, spurious correlation describes a scenario in which two variables seem to be associated, but their relationship is either accidental or influenced by an external factor~\cite{ye2024spurious}. Such situations can result in misleading or erroneous interpretations of data and models~\cite{ye2024spurious}. Supervised learning algorithms are vulnerable to spurious correlations because classifiers often tend to learn any signal in the dataset that maximizes accuracy, even those that may appear incomprehensible to humans~\cite{ilyas2019adversarial}. Spurious correlations pose significant challenges for ML in cybersecurity because existing ones may lead ML models to learn patterns in the data unrelated to the security problem, thereby creating shortcuts for classifying classes~\cite{quiring2022and}. 
Over the past few years, several approaches (e.g., invariant learning~\cite{arjovsky2019invariant} and group robustness~\cite{yang2022chroma}) have been proposed to mitigate the impact of spurious correlations in ML. Domain adaptation stands out as one of the approaches aimed at aligning the distribution between source and target domains to address spurious correlations. This approach focuses on transferring knowledge learned from the source domain to the target domain, thereby aiding the model in better generalizing to new data and reducing its dependence on spurious correlations~\cite{li2023invariant}. In our study, we concentrate on domain adaptation by suggesting a robust feature space, as a high-quality feature representation is vital for the success of domain adaptation~\cite{ben2006analysis}.

\begin{figure}[t]
    \centering
    \includegraphics[width=0.9\columnwidth]{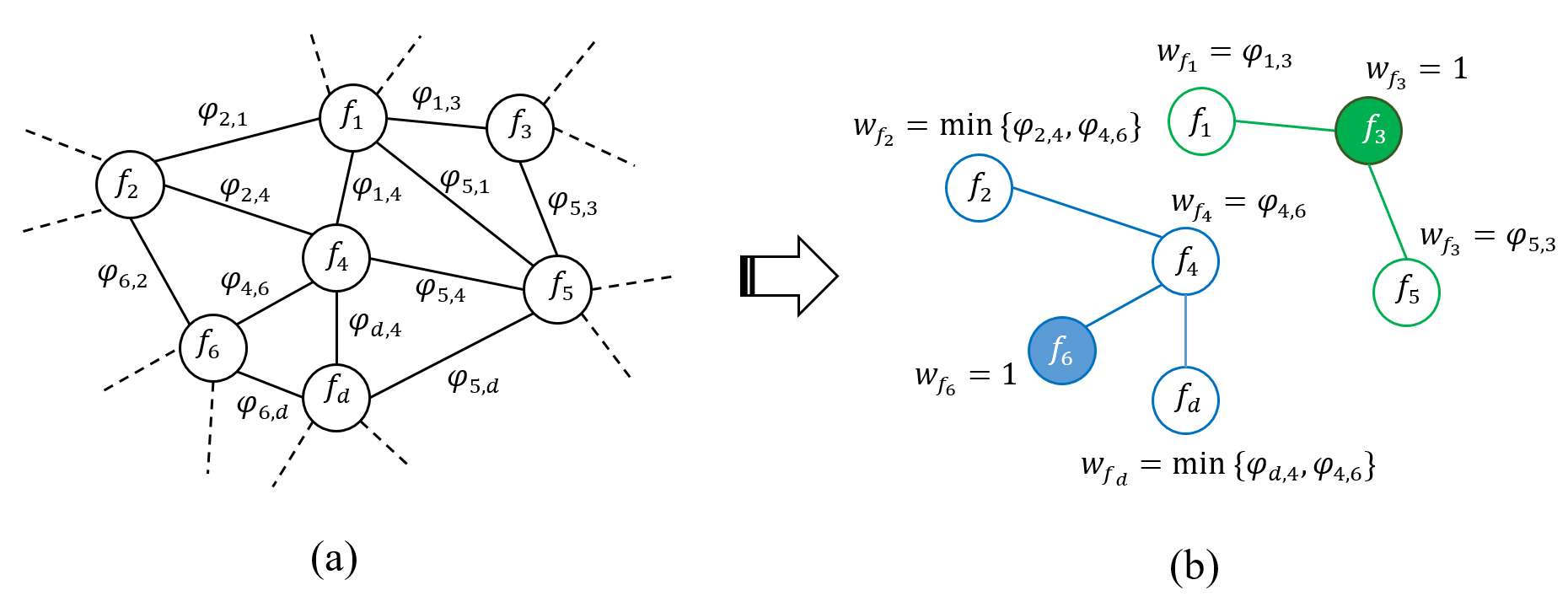}
    \caption{An example of the OPF process constructed to capture feature dependencies. (a) shows the completed weighted graph $\set G$, where node $f_i$ represents the feature $f_i$ and $\varphi_{i,j}$ represents the correlation between $f_i$ and $f_j$. (b) the final OPF comprises two OPTs derived from $\set G$. The colored nodes signify primary features, and $w_{f_i}$ denotes the path cost from $f_i$ to its relevant primary features.
    }
    \label{fig:opf}    
\end{figure}

\subsection{Domain Constraints in the Feature Space}
\label{subsec:problem-space_constraints_interpretation}
Generally, evasion attacks must account for domain constraints to generate realizable AEs. In the problem space, the domain constraints consist of available transformations, preserved semantics, robustness to preprocessing, and plausibility, as formalized by Pierazzi et al.~\cite{b17}. These constraints signify the properties of malware apps that must be maintained after manipulation in the problem space.
However, domain constraints appear differently in the feature space. During the past years, numerous studies in diverse contexts have demonstrated that feature dependencies serve as indicators of domain constraints within the feature space~\cite{b39,b51,b68,b40,b70}. In this study, we apply the idea proposed by~\cite{b40} to infer domain constraints based on feature dependencies. This approach suggests that the relationship between primary and secondary features can model domain constraints. Primary features are those that limit the range of permissible values for other features, whereas secondary features do not impose any such limitations. According to~\cite{b40}, primary features can be identified based on data observations by indicating the features that are most correlated with others. Here, we not only employ correlation to specify the primary features but also adapt Optimum-path Forest (OPF), as used in~\cite{bostani2022domain}, to determine the secondary features. Using the primary features as a starting point, we apply OPF to partition the remaining features into different groups, ensuring that each cluster includes relevant features. (i.e., primary and secondary features). In fact, we partition $\set F$, the feature set characterizing $\set X$, into distinct clusters denoted as $\set A_i$. This clustering strategy aims to identify primary and secondary features by grouping them together. As shown in Figure~\ref{fig:opf}, the technique involves constructing a complete weighted graph, denoted as $\set G=(\set V, \set E)$, where $\set G = \set F$ and $\set E = \set F \times \set F$ encompassing edges that connect each feature pair $(f_a, f_b)$, with weights determined by a correlation coefficient. Following the graph construction, the next step involves partitioning $\set G$ into various clusters such as $\set A_i$ where it constitutes a subset of all features within the feature space, i.e., $\set A_i \subset \set F$. It is noted that the OPF algorithm treats each cluster as an Optimum-path Tree (OPT), wherein each feature has an optimal path to the OPT's prototype, which is a primary feature in our case. The path cost is the minimum weight of the edges along the path. For more details about primary and secondary features, refer to~\cite{b40}, and for information on OPF, see~\cite{bostani2022domain}.

%% file: 03-our_proposed_defense.tex
\section{Our Proposed Defense}
\label{section:proposed_defense}
This section illustrates how domain constraints, specified by feature dependencies, have been utilized to propose a robust feature space. The proposed defense aims to enhance the robustness of ML-based AMD against realistic attacks by reducing the impact of spurious correlations. To this end, we first formulate a new domain adaptation approach, which leverages the feature dependencies of the source domain to build a resilient feature space. Subsequently, we delve into an elaborate discussion demonstrating the process of constructing our proposed robust feature space.

\subsection{Formulation of the Problem} 
Suppose $\mathcal{X}$ denotes the initial feature space, and $\set T$ denotes the source domain (i.e., training set), and $\mathcal U$ implies the target domain (i.e., unseen samples). Moreover, let $\mathcal{Y}$, ${\mathcal{D}_\mathcal{T}}^\mathcal{X}$, and ${\mathcal{D}_\mathcal{U}}^\mathcal{X}$ represent the label space and the data distributions of $\set T$ and $\set U$ based on $\mathcal{X}$, respectively. Given a labeled source dataset $\{ (\mathbf{x}_i, y_i) \}_{i=1}^{n}$ drawn from ${\mathcal{D}_\mathcal{T}}^\mathcal{X}$, the goal is to build a robust feature space $\mathcal{H}$ that captures the dependencies observed in $\mathcal{X}$ and can be effectively used for classification in both $\mathcal{T}$ and $\mathcal{U}$. We can mathematically express this as follows:
\begin{enumerate}
    \item Construct a feature transformation function $\lambda: \mathcal{X} \rightarrow \mathcal{H}$ that maps the initial feature space $\mathcal{X}$ to the robust feature space $\mathcal{H}$. This function is designed to capture the relevant feature dependencies observed in $\mathcal{X}$ considering the source domain and build a new feature representation based on them. Indeed, the transformation function tends to construct $\mathcal{H}$ wherein ${\mathcal{D}_\mathcal{T}}^\mathcal{H}$ and ${\mathcal{D}_\mathcal{U}}^\mathcal{H}$ are more align compared to ${\mathcal{D}_\mathcal{T}}^\mathcal{X}$ and ${\mathcal{D}_\mathcal{U}}^\mathcal{X}$.
    \item Train a classifier $f: \mathcal{H} \rightarrow \mathcal{Y}$ using the transformed features $\mathcal{H}$. This classifier should be capable of making accurate predictions based on robust feature representations. The objective can be formulated as minimizing a loss function $\mathcal{J}$ over the labeled source dataset:
    \[
    \min_{\lambda, f} \sum_{i=1}^{n} \mathcal{J}(f(\lambda(\mathbf{x}_i)), y_i)
    \]    
    where $\mathcal{J}$ represents the classification loss function, and $f(\lambda(\mathbf{x}_i))$ denotes the predicted label for the $i$-th source domain sample after feature transformation.
\end{enumerate}

\subsection{Robust Feature Space}
\label{section:robust_feature_space}
The primary objective of our proposed robust feature space is to mitigate the impact of adversarial perturbations on misclassification by aligning the distributions of malware samples seen during training and AEs. Leveraging the proposed robust feature space adapts the ML-based malware detection, trained on training samples, to perform well on samples with unseen distribution, especially AEs. Generally, the adversarial perturbations used in realistic attacks consist of redundant codes that are unrelated to the malicious functionality of the original malware apps. Therefore, we anticipate that their adversarial effect can be diminished if our ML model can learn authentic malicious patterns rather than shortcut patterns (i.e., spurious correlations~\cite{quiring2022and}) unrelated to malware detection. To achieve this, we propose a new transformation function (i.e., $\lambda:\set X\rightarrow \set H$) that maps each $x \in \set X$ to an $h \in \set H$ based on feature-space domain constraints. ML used in AMD operates with feature representations of apps in $\set H$, which implicitly abstract domain constraints, instead of $\set X$ in both the training and inference phases. We expect that training our ML-based detector on $\set H$ gives the detector more chances to learn generic attack patterns to distinguish malicious and benign apps because $\set H$ is characterized based on the pre-identified patterns (i.e., a collection of feature groups, where each includes the interdependent features) that represent domain constraints. Indeed, training an ML model on $\set H$ will cause the detector to rely on groups of features in $\set X$ rather than individual features, potentially introducing bias in classification.

Suppose $\Lambda = \{ \set A_1, \set A_2, ..., \set A_m\}$ represents all feature dependency clusters identified by utilizing the method discussed in Section~\ref{subsec:problem-space_constraints_interpretation}, 
where $\set A_i \subset \set F$ denotes the feature set of the $i$-th cluster in $\Lambda$. Additionally, let $\set L$ denote the dimensions (i.e., features) of the feature space $\set H$. As depicted in Figure~\ref{fig:RobustFeatureSpace}, within the proposed new feature space $\set H$, each $\set A_i$ is associated with a feature $l_i \in \set L$ serving as the representative of all features in $\set A_i$ to ensure the detection model's accuracy on legitimate samples. This representativeness is necessary since $\set A_i$ encompasses interdependent features with similar information about the target class, making a single feature in $\set L$ sufficient to represent these relevant features. To determine whether a feature $l_i \in \set L$ should appear in $\set H$ for a sample $x \in \set X$, we utilize an activation function based on the sigmoid. This function can transfer the influence of input features in $\set A_i$ to the output feature $l_i$ while uniformly increasing the probability of a feature appearing in the output as the number of input features rises. Furthermore, the function ensures that changing a feature $f_j \in \set A_i$ cannot simply alter $l_i$ due to our aim to increase the evasion costs for adversaries. It is important to note that the sigmoid function, being a monotonic function, adjusts $l_i$ based on the features in $\set A_i$. As the sigmoid function exhibits an \textit{S-Shaped Curve}~\cite{sandberg2021sigmoids}, we anticipate that it can help smooth out the severe impact of adversarial perturbations within $\set A_i$ on $l_i$. This is because, when there are large adversarial fluctuations or spikes in the input feature values in $\set A_i$, the output will show a more gradual change due to the sigmoid's property of saturating large values. According to the proposed transformation method, the value of feature $l_i$ in $h = \lambda(x)$ is computed as follows:
\begin{figure}[t]
    \centering
    \includegraphics[width=0.4\columnwidth]{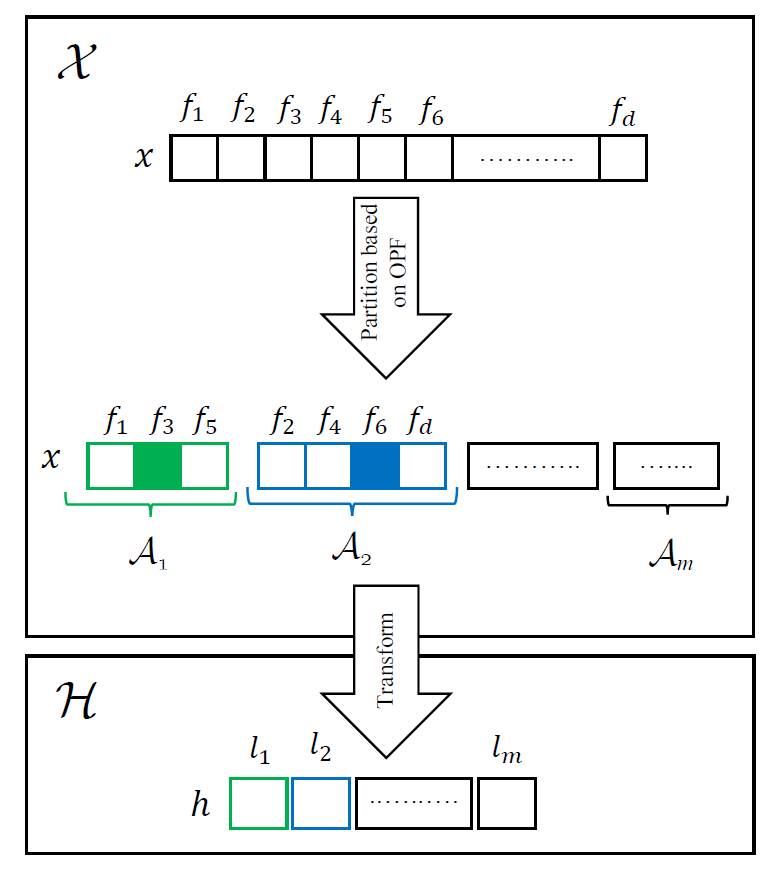}
    \caption{Overview of our method for applying domain constraints to construct a robust feature space.}
    \label{fig:RobustFeatureSpace}    
\end{figure}
\begin{equation}
l_i =
    \begin{cases}
      1 & \text{$\sigma(s=\sum_{\forall f_j \in \set A_i} w_{j} \cdot x_{j}) > \theta$}\\
      0 & \text{otherwise},
    \end{cases}   
   \label{eq_activation_function}
\end{equation}
where $\sigma(\cdot)=\frac{1}{1+e^{-s}}$ represents the sigmoid function, $w_{j}$ stands for the weight of feature $f_j$ in $\set A_i$, $x_{j}$ denotes the value of feature $f_j$ in $x$, and $\theta$ signifies a threshold for activating $l_i$ in $\set H$. It's important to note that not every feature holds the same level of importance (e.g., some features may arise due to noise). Thus, we aim to account for the greater impact of features that might contribute more to the detection task. Consequently, as shown in equation~(\ref{eq_activation_function}), we take into consideration the weights of features since they can indicate the importance of features in $\set A_i$. In the proposed transformation function, we consider the path cost of each feature in the constructed OPF as the weight for the features, since it serves as a measure demonstrating the relevance of a certain feature to its respective cluster. Moreover, in equation~(\ref{eq_activation_function}), $s\geq 0$ since $w_j$ possesses a positive value. Hence, $0.5\leq\sigma \leq 1$ in our context, implying that $\theta$ should be chosen from the interval $(0.5,1)$. Generally, opting for a moderate threshold seems preferable because setting a very low threshold may result in the output feature appearing even with minor perturbations in the input features. This occurs because triggering more features in $\set A_i$ leads to a larger $\sigma$, potentially causing $l_i = 1$ if $\theta$ is small. On the other hand, a very high threshold hinders the transfer of the input features' effect to the output. In essence, although a high threshold might enhance the model's robustness by reducing the impact of adversarial perturbations on features in $\set A_i$, it could decrease the model's accuracy on legitimate samples.

%% file: 04-experiments.tex
\section{Experiments}
\label{section:experiments}
In this section, we conduct empirical assessments to gauge the effectiveness of the proposed defense mechanism against various realistic evasion attacks. All experiments were conducted on a Debian Linux workstation equipped with an Intel(R) Core(TM) i7-4770K CPU running at 3.50 GHz and 32 GB of RAM.

\subsection{Experimental Setup}
\noindent\textbf{Dataset.} This study utilizes an available dataset~\cite{b17} comprising approximately $170K$ Android apps sourced from AndroZoo~\cite{b67}. An app within this dataset is classified as benign if no VirusTotal Antiviruses (AVs) detect it, while it is deemed malware if it is flagged by four or more AVs. Our training set comprises $50K$ randomly selected samples, with $30K$ samples allocated for the test set for evaluating Android malware detectors. The training set consists of $45K$ clean samples and $5K$ malware samples, whereas the test set comprises $25K$ clean samples and $5K$ malware samples. All samples are encoded based on the DREBIN~\cite{b9} feature space before being processed by the malware detectors. Given that DREBIN encompasses a vast but sparse feature space, we select the $10K$ most frequently occurring features, as recommended by prior research~\cite{b19,b20}.
To evaluate the adversarial robustness of various Android malware detectors, we employ $1K$ malware samples as outlined in~\cite{b48} to generate AEs.

\noindent\textbf{Threat Models and Attacks.}
Adversarial attacks can be analyzed based on their objectives, knowledge, and capabilities. The adversary's objective is to cause misclassification in AMD, resulting in the classification of the adversarial (malware) examples as benign ones. Moreover, the adversary's knowledge of the target model, including its training data, feature space, and parameters, ranges from perfect (PK) to limited (LK) or zero (ZK). In PK, LK, and ZK attacks, the target model is perceived as a white-box, gray-box, and black-box model, respectively. Finally, the adversary's capability enables the generation of AEs either within the feature space, by altering feature representations of Android malware apps, or within the problem space, through a sequence of transformations applied to Android malware apps~\cite{b17}.

This paper explores two realistic problem-space attacks, namely \textit{PK-Greedy}~\cite{b17} and \textit{EvadeDroid}~\cite{b48}, to evaluate the adversarial robustness of malware detectors discussed in \S\ref{section:detectors_with_different_feature_spaces}. PK-Greedy and EvadeDroid transform Android apps into adversarial instances by targeting white-box and black-box malware detectors, respectively. The details of these attacks are described as follows.
\begin{itemize}
\item\textbf{PK-Greedy}~\cite{b17} generates problem-space realizable AEs by applying effective transformations (i.e., code snippets called gadgets extracted from donor apps) specified by feature-space perturbations on the target model. This attack adds not only \textit{primary} features to bypass malware detection but also \textit{side-effect} features to meet the domain constraints. The attack was originally tested in the PK setting, but here we also test it in an LK setting where the AEs transfer from a surrogate model to a target model. To utilize PK-Greedy in PK settings when the feature space is $\set H$, we must adapt this attack to target the ML model trained on the robust feature space, as it was initially designed for models trained on the DREBIN feature space. Therefore, PK-Greedy is an adaptive one aware of our proposed transformation function. Specifically, during the attacking phase, PK-Greedy identifies the most adversarially sensitive features in $\set L$ based on the model trained on $\set H$ where $\set L$ is the feature set characterizing $\set H$. Then, for each identified feature $l_i \in \set L$, it finds a transformation wherein its triggered features (i.e., the DREBIN features that can appear in an app after applying the transformation) have a significant overlap with the features in $\set A_i$ (i.e., the cluster in the DREBIN feature set corresponding to $l_1 \in \set L$), and then applies it to the app.

\item\textbf{EvadeDroid}~\cite{b48} generates problem-space realizable AEs through a sequence of transformations by querying the target model in a ZK setting. This adversarial attack involves the initial collection of problem-space transformations by extracting code snippets containing API calls from benign apps found in the wild, resembling malware apps. Subsequently, random search is employed to select and apply transformations that induce the malware app to exhibit similarities to benign apps. In addition to the original ZK setting, here we also consider a more restricted setting where EvadeDroid is only allowed to query a surrogate model and then transfer the AEs to the target model. Specifically, we set the query budget $Q = 10$, and $\alpha = 50\%$ (i.e., the percentage of the relative increase in the size of a malware sample after manipulation).
\end{itemize}

\noindent\textbf{Evaluation Metrics.}
To assess the malware detectors, we test both their clean performance and robustness.
For clean performance, we compute Clean Accuracy, True Positive Rate (TPR), and False Positive Rate (FPR) on benign and malware samples.
For robustness, we compute Robust Accuracy on adversarial malware examples. Additionally, we include the average number of added features needed to achieve successful AEs.

\subsection{Evaluation of Proposed Defense}
\label{section:detectors_with_different_feature_spaces}
This section aims to evaluate our robust feature space introduced in \S\ref{section:proposed_defense}. We consider the following four ML-based Android malware detectors: 
\begin{itemize}
    \item \textbf{DREBIN-Original}~\cite{b9}, a well-known Android malware detector that is based on the linear Support Vector Machine (SVM). It is trained with the original DREBIN feature space.
    \item \textbf{Sec-SVM}~\cite{b19} is the secure version of DREBIN-Original for strengthening the robustness of linear SVM against AEs. Sec-SVM relies on more features, and this increases the evasion cost. Essentially, the goal of Sec-SVM is to enhance the robustness of a linear SVM by ensuring that it assigns weight more evenly across all features used in the model. This approach inherently makes generating AEs more challenging for an attacker, as it needs to alter more features to bypass malware detection. 
    \item \textbf{DREBIN-Robust} is our robust DREBIN detector trained with our new robust feature space.
    \item \textbf{DREBIN-FeatureSelect} resembles DREBIN-Original but is trained with a lower-dimensional feature space to enhance the stability of models against noise~\cite{b63}. Feature selection aims to eliminate redundant or irrelevant features, which can be misused by adversarial perturbations, and thus improve the robustness of an ML model. We utilize Linear SVC to identify the 500 most influential features, resulting in clean accuracy comparable to DREBIN-Robust. 
\end{itemize}

\begin{table}[!t]
\caption{
The Training Time (s), clean performance metrics (including TPR (\%), FPR (\%), Clean Acc (\%)), and Robust Acc (\%) for various DREBIN detectors.}
Realizable AEs are transferred from DREBIN-Original. ``*'' means both the surrogate and target models are DREBIN-Original.
\begin{center}
\begin{tabular}{l|r|r|r|r|r|r}
\hline
{\textbf{Model}}&{\textbf{\Tstrut\shortstack{Training\\Time}}} & \textbf{TPR} & \textbf{FPR} & {\textbf{\shortstack{Clean\\ Acc}}}&{\textbf{\shortstack{Robust Acc\\  (PK-Greedy})}}&{\textbf{\shortstack{Robust Acc\\ (EvadeDroid)}}}\\
\Xhline{0.75pt}
DREBIN-Original & 8.2 & 87.2 & 1.4 & 96.7 & *0.0& *26.8\\
\Xhline{0.25pt}
Sec-SVM & 25.4 & 77.0 & 1.0 & 95.3 & 97.6 & 72.1
\\
\Xhline{0.25pt}
DREBIN-FeatureSelect  & 1.3 & 79.5 & 1.3 & 95.5 & 30.7 & 49.8\\
\Xhline{0.25pt}
DREBIN-Robust (ours)  & 1.7 & 77.5 & 1.3 & 95.1 & 97.9 & 94.6\\
\hline
\end{tabular}
\label{table:performance_of_svms}
\end{center}
\end{table}

\begin{figure}[!b]
    \centering
    \includegraphics[width=8cm]{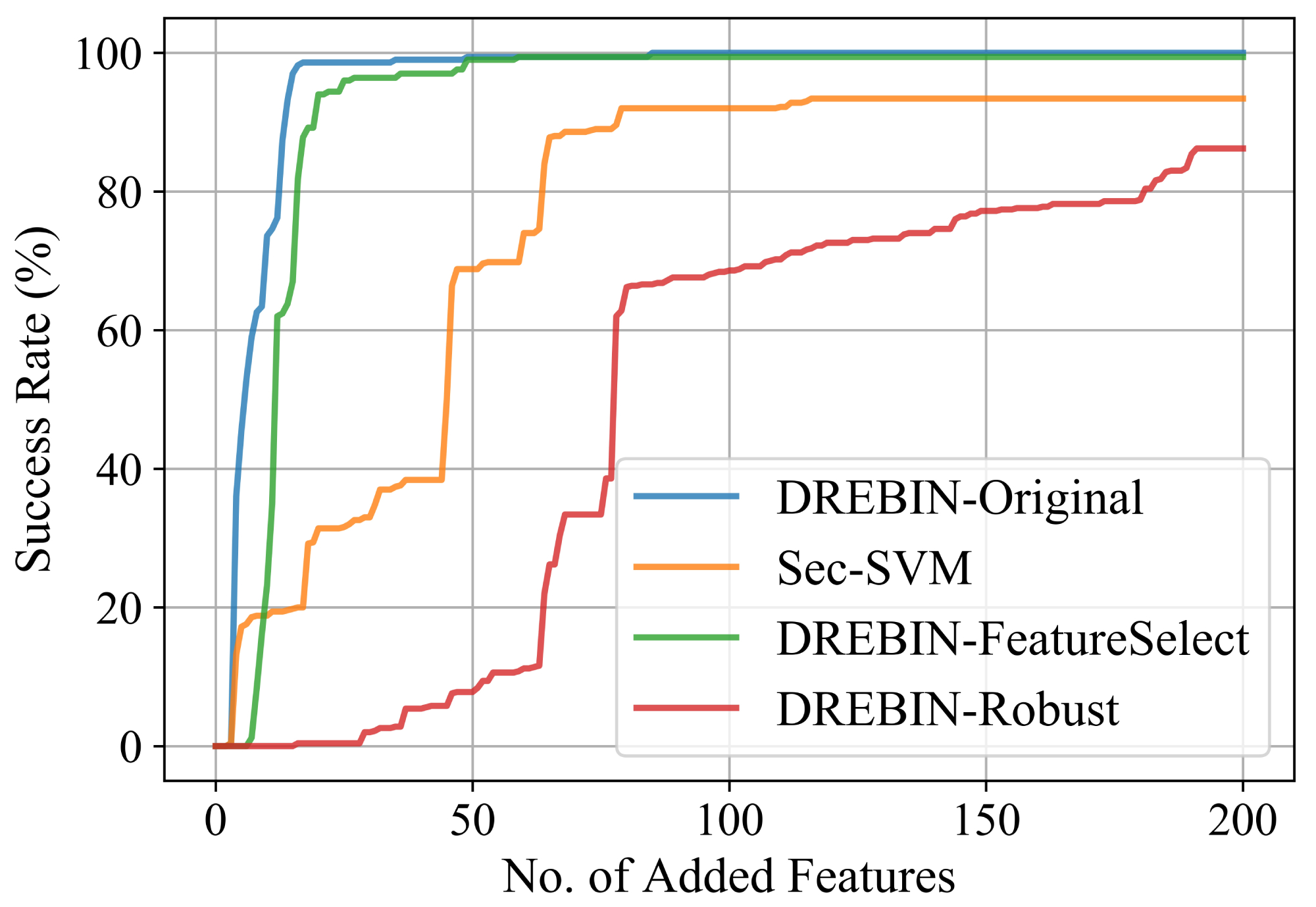}
    \caption{The evasion success rates of PK-Greedy against different DREBIN detectors when varying the number of added features.}
    \label{fig:ers_against_l1_norms}
\end{figure}

In the first experiment, we assess the robustness of malware detectors against transferable AEs generated by PK-Greedy and EvadeDroid, using DREBIN-Original as the surrogate model. It's ensured that the AEs are generated from malware samples correctly detected by all four malware detectors, and the results are computed based on the successful AEs for DREBIN-Original. As depicted in Table~\ref{table:performance_of_svms}, all defenses achieve similar clean performance in terms of TPR, FPR, and clean accuracy; however, the training time of our proposed defense is substantially shorter than that of Sec-SVM and even DREBIN-Original. It is important to note that, despite the significant improvement in training time, our technique incurs some overhead due to the creation of the robust feature space and the transformation of training samples into this space. However, this overhead is minimal compared to the feature engineering required during preprocessing to extract and represent features from apps in the initial feature space. Concerning robustness, although DREBIN-FeatureSelect demonstrates notable robustness compared to DREBIN-Original due to feature selection, both DREBIN-Robust and Sec-SVM exhibit significantly higher robustness. Additionally, our DREBIN-Robust outperforms Sec-SVM,
especially for EvadeDroid.

\begin{table}[!t]
\centering
\caption{The robustness of different DREBIN detectors against realizable AEs that are directly generated on the corresponding models in terms of Robust Acc~(\%) and Number of Features (NoF).}
\begin{tabular}{l|r|r|r|r}
\Xhline{0.25pt}
\multirow{2}{*}{\textbf{ Model}} & \multicolumn{2}{c}{\textbf{PK-Greedy}}& \multicolumn{2}{c}{\textbf{EvadeDroid}} \\
\cline{2-5} 
  & \textbf{\shortstack{ Robust Acc}}
  & \textbf{ NoF}
  & \textbf{\shortstack{ Robust Acc}}
   & \textbf{ NoF}\\
\Xhline{0.75pt}
 DREBIN-Original &  0.0 &   9.2 &  26.8 &  66.5 \\
 \Xhline{0.25pt}
Sec-SVM &  6.6 &   37.9 &  31.9 &  77.5 \\
\Xhline{0.25pt}
DREBIN-FeatureSelect & 0.6 &  19.6 & 45.9 & 62.1 \\
\Xhline{0.25pt}
 DREBIN-Robust (ours) &  13.8 &  86.1  &  87.0 &  56.7\\
\Xhline{0.25pt}
\end{tabular}
\label{table:robust-feature-representation}
\end{table}

We further examine a more challenging scenario where AEs are directly generated on the target model. Due to the time-consuming nature of generating problem-space AEs across different detectors, we limit our test to 500 malware samples. Moreover, to ensure rigorous evaluation of detectors in worst-case scenarios, PK-Greedy operates in the PK setting, especially in attacking the ML model trained on the robust feature space (i.e., DREBIN-Robust). As illustrated in Table~\ref{table:robust-feature-representation}, the superiority of our DREBIN-Robust over Sec-SVM remains evident, particularly in defense against EvadeDroid. Indeed, DREBIN-Robust demonstrates superior performance compared to Sec-SVM against both attacks, with a 7.2\% improvement against PK-Greedy and a 55.05\% improvement against EvadeDroid. Additionally, in defense against PK-Greedy, DREBIN-Robust escalates the evasion cost by necessitating the adversary to modify significantly more features to achieve success. Figure~\ref{fig:ers_against_l1_norms} further validates the consistently superior performance of our DREBIN-Robust across varying numbers of added features. Furthermore, we observe rapid convergence of the evasion rates, suggesting that increasing the number of added features scarcely improves PK-Greedy against our DREBIN-Robust. It should be noted that although our defense in DREBIN-Robust substantially improves robustness against both realistic attacks, the success rate of PK-Greedy is significantly higher than that of EvadeDroid--specifically, 73.16\%--because PK-Greedy is an adaptive attack that operates in PK settings, whereas EvadeDroid targets our proposed detector in ZK settings.

\begin{figure}[!b]
    \centering
    \includegraphics[width=0.9\columnwidth]{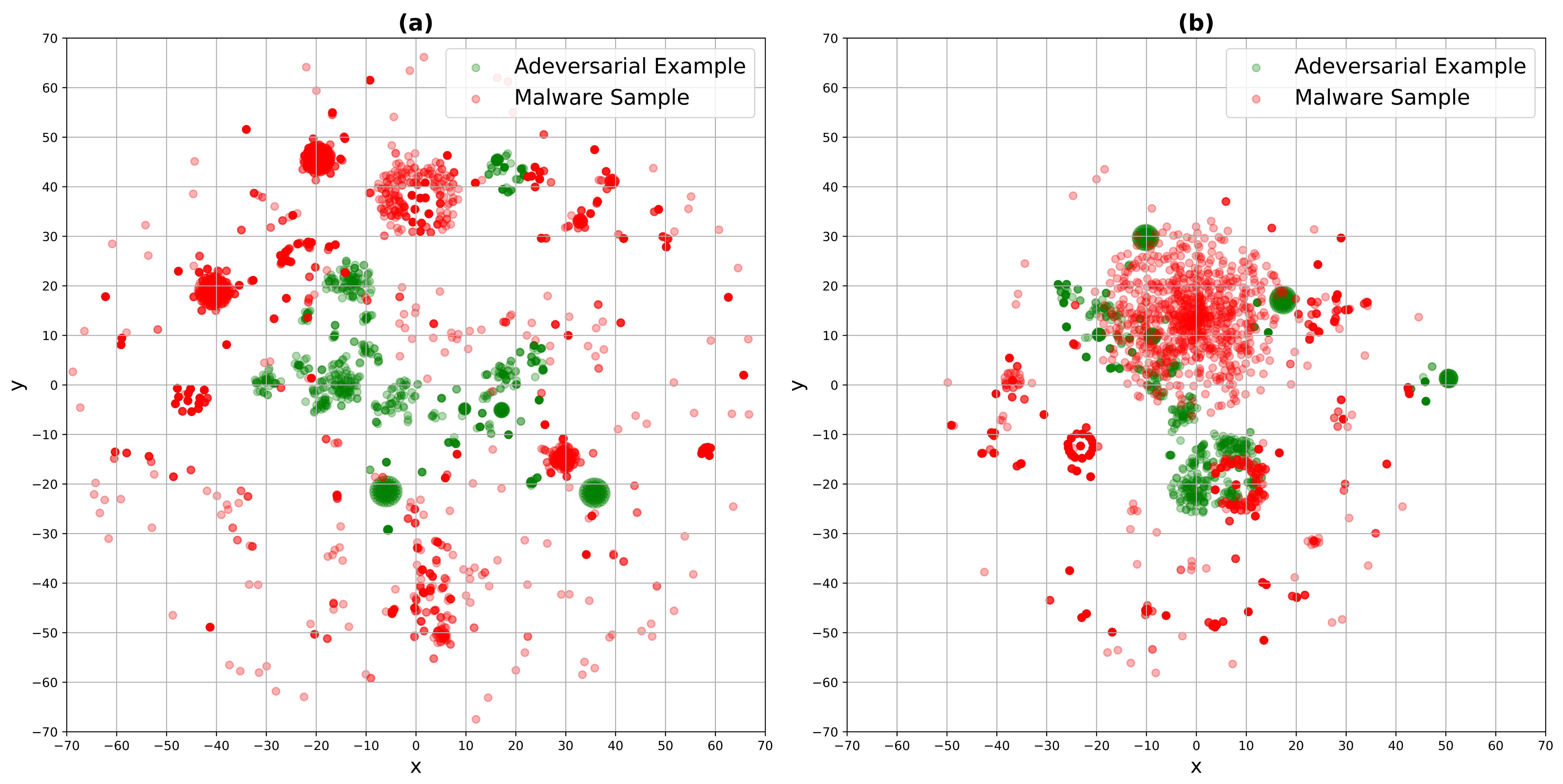}
    \caption{t-SNE visualization of malware and adversarial malware samples in (a) the feature space $\mathcal{X}$ and (b) our robust feature space $\mathcal{H}$.}
    \label{fig:tsne}
\end{figure}

Note that in Table~\ref{table:robust-feature-representation}, the fact that the average number of added features required by EvadeDroid for bypassing DREBIN-Robust is lower than bypassing the other detectors is due to the property of EvadeDroid. Specifically, in EvadeDroid, a transformation is applied to a malware app only if it can increase the chance of generating successful AEs. This leads to the difference between the number of transformations applied to the detectors (e.g., $1.00$ for DREBIN-Robust vs. $2.54$ for Sec-SVM). This difference indicates that most transformations are not good enough for attacking DREBIN-Robust, and consequently leads to a difference in the number of added features.

\subsection{Discussion}
Our empirical investigation highlights the resilience of the proposed robust feature space against realizable AEs generated by various realistic evasion attacks. To further analyze this observation, Figure~\ref{fig:tsne} displays the t-SNE visualization of malware samples from the training set and AEs from the test set. The visualization is based on the top-100 important features selected using Linear SVC, both in the original feature space and our proposed robust feature space. It is noteworthy that AEs are generated by targeting DREBIN-Original with PK-Greedy. In the proposed robust feature space, the visualization demonstrates a closer alignment between the distribution of malware samples and AEs compared to those observed in the initial feature space. Consequently, this contributes to the enhanced adversarial robustness observed in DREBIN-Robust trained on $\mathcal{H}$ compared to DREBIN-Original trained on $\mathcal{X}$, as demonstrated in the results presented in Table~\ref{table:performance_of_svms}. 

Moreover, to grasp the extent of our approach's efficacy in countering potentially misleading correlations learned by models, we delve into the operational mechanism of DREBIN, (i.e., a linear SVM-based detector). The detection model relies on a score function, derived from the inner product of the model's parameters (i.e., learned weights $\vec{W}$) and a feature vector representing an app $z$. Specifically, this function is denoted as $f(z) = \langle \phi(z), \vec{W} \rangle$ when the model is trained on $\mathcal{X}$, and as $f(z) = \langle \lambda(\phi(z)), \vec{W} \rangle$ when trained on $\mathcal{H}$. A sample is classified as benign if $f < 0$, and conversely if $f > 0$. This suggests that features with high negative weights play a pivotal role in classifying a sample as benign. Adversaries target altering the features relevant to negative weights as it increases the chance of deceiving malware detectors. The features that are important for the classifier operating on $\set X$ might be biased features; however, $\set H$ aims to diminish classifier bias by assisting it in relying on sets of features that contribute more to the behavior of the apps, rather than on individual features. For instance, as shown in Figure~\ref{lst:features}, $f_1 \in \set X$, the feature with the most negative weight in the linear SVM trained on $\mathcal{X}$,  can potentially represent a shortcut feature that might be important due to biased data, while $l_1 \in \set H$, the feature with the most negative weight in the linear SVM trained on $\mathcal{H}$, seems relevant to functionality.

\begin{figure}[b]
\centering
\begin{lstlisting}[basicstyle=\fontsize{6}{7}\selectfont\ttfamily]
{
    "f1": [    
      'URLs::https://play.google.com/store/apps/'    
    ],

    "l1": [
      'Activities::xmlparser.GiftActivity',
      'URLs::http://jgpre.alibaba.inc.com/',
      'URLs::http://jg.daily.taobao.net/',
      'URLs::http://jg.alibaba.inc.com/',   
      'Activities::xmlparser.SplashScreenActivity',   
      'Activities::xmlparser.PrivacyActivity', 
      'S_and_P::android.permission.BIND_REMOTEVIEWS'
    ]
}
\end{lstlisting}
\caption{The details of features $f_1 \in \set X$ and $l_1 \in \set H$.}
\label{lst:features}
\end{figure}

We conduct a further evaluation to ascertain whether DREBIN-Robust exhibits superior resilience compared to DREBIN-Original in learning spurious correlations. To ensure bias in a feature, we must identify a feature that not only seems biased but also is absent in some of the malware samples in our test set, allowing us to observe the effects of adding it to the malware samples. Note that a feature appears biased when it is prevalent in the majority of benign samples but is present in the minority of malware samples within the training set. Among features with negative weights, feature $f_{44}$ (i.e., \texttt{android.permission.INTERNET}) stands out as the first one that not only appears biased but is also absent in some of the malware samples in the test set. Adding this feature to malware samples of the test set that lack $f_{44}$, drops DREBIN-Original's robustness from 96.4\% to 86.3\%. This demonstrates $f_{44}$'s bias, resulting in spurious correlations in DREBIN-Original. This is because the feature is not effective in distinguishing between benign and malware samples since it can potentially be present in both types of samples. However, DREBIN-Robust remains unaffected by this misleading correlation, demonstrating its resilience against modifications to feature $f_{44}$.

\noindent\textbf{Ethical Considerations.} Since our proposed defense strategy is designed to enhance cybersecurity and mitigate adversarial attacks, rather than facilitate malicious activities, ethical concerns are minimal. However, we stress that our defense mechanisms should be used responsibly and primarily as a baseline for research purposes.

%% file: 05-related_work.tex
\section{Related Work}
\label{section:related-work}

Despite numerous efforts aimed at enhancing the adversarial robustness of ML-based AMD against evasion attacks (e.g., ~\cite{b1,b20,b21,b28,b29,b30}), few studies have primarily explored the impact of features on enhancing adversarial robustness. To this end, Demontis et al.~\cite{b19} introduced Sec-SVM which trains a linear SVM with a more uniform distribution of feature weights. This study ensured that the linear model learned feature weights more evenly by applying box constraints on weights within the standard optimization problem used in the linear SVM. Their proposed Sec-SVM relies on a larger number of features for classification, thereby enhancing adversarial robustness, as attackers would need to perform significantly more meticulous manipulations to generate AEs. Chen et al.~\cite{chen2017securedroid} introduced SecureDroid, a defense strategy employing ensemble learning, coupled with a novel feature selection technique, to bolster classifier resilience against evasion attacks. The proposed method highlighted the importance of individual features, considering both their contribution to classification and their vulnerability to manipulation by attackers. In other words, the paper argues that the features with greater significance in classification and lower manipulation cost are interesting features for attackers. Specifically, the proposed method reduced the presence of these features by altering the training set. This resulted in a more uniform distribution of feature importance, compelling attackers to manipulate a larger array of features to bypass detection. Yang et al.~\cite{b30} investigated weight bounding, similar to~\cite{b19}. They constrained the weights of the linear classifier on a few dominant features to achieve more evenly distributed feature weights. They determined the dominant features by noting that adversaries could generate evasive malware variants with minimal mutations on certain features identified as dominant, compared to the extensive mutations required on other non-dominant features. Chen et al.~\cite{b44} introduced a gradient masking method that converts the binary feature space into continuous probabilities, encoding the distribution for both benign and malicious instances. The paper argues that when the binary feature space is transformed into a continuous space, the gradient of feature addition or removal accessible to attackers may be significantly reduced. As a result, attackers cannot easily bypass detection. This technique also enables the classifier to strike an optimal balance between security and accuracy by utilizing a softmax function with an adversarial parameter.

%% file: 06-limitations_and_future_work.tex
\section{Limitations and Future Work}
\label{sec:limitations_and_future_work}
While our experiments on DREBIN convincingly showcase the effectiveness of our proposed defense against realistic evasion attacks, it remains imperative to extend our evaluation to encompass a broader array of malware detection systems. By doing so, we can ascertain the generalizability of our approach across different detectors.
Furthermore, since the technique for capturing domain constraints is data-driven, it is essential to periodically update the feature space before regular retraining to keep ML models effective. This is crucial for maintaining the performance of malware classifiers, especially against evolving zero-day malware with varying feature dependencies. In addition, as elucidated in \S\ref{section:robust_feature_space}, the selection of an appropriate threshold is pivotal for ensuring the efficacy of our defense. Hence, future inquiries should delve into the impact of different threshold values on both the clean and robust accuracies of our defense mechanism.

%% file: 07-conclusion.tex
\section{Conclusion}
\label{section:conclusion}
In this study, we introduce a novel defense mechanism based on domain adaptation to enhance the adversarial robustness of ML-based AMD. The proposed method aims to enhance the reliable generalizability of AMD against adversarial examples by mitigating spurious correlations misused by evasion attacks. Our approach leverages domain constraints to establish a robust feature space, enabling ML models to learn genuine malicious patterns of Android malware. Experimental results on DREBIN, a well-known AMD, demonstrate significant improvements over the state-of-the-art defense Sec-SVM, particularly against realistic evasion attacks.